\documentclass[aps,prb,reprint,twocolumn,floatfix]{revtex4}
\usepackage{graphicx}
\usepackage{amsmath,amssymb}
\usepackage{url}

\begin{document}

\title{Osmosis in a minimal model system}
\author{Thomas W. Lion}
\email{tom.wlion@gmail.com}
\author{Rosalind J. Allen}
\affiliation{SUPA, School of Physics and Astronomy, University of Edinburgh, James Clerk Maxwell Building, The King's Buildings, Mayfield Road, Edinburgh, EH9 3JZ, UK}

\begin{abstract}
Osmosis is one of the most important  physical phenomena in living and soft matter systems. While the thermodynamics of osmosis is well understood, the underlying  microscopic dynamical mechanisms remain the subject of discussion. Unraveling these mechanisms is a prerequisite for  understanding osmosis in non-equilibrium systems. Here, we investigate the microscopic basis of osmosis, in a system at equilibrium, using  molecular dynamics simulations of a minimal model in which repulsive solute and solvent particles differ only in their interactions with an external potential. For this system, we can derive a simple virial-like relation for the osmotic pressure. Our simulations support an intuitive  picture in which the solvent concentration gradient, at osmotic equilibrium, arises from the balance between an outward force, caused by the increased total density in the solution, and an inward diffusive flux caused by the decreased solvent density in the solution. While more complex effects  may occur in other osmotic systems, our results suggest that they are not required for a minimal picture of the dynamic mechanisms underlying osmosis.
\end{abstract}

\maketitle

\section{Introduction}

Osmosis is the physical phenomenon in which a concentration difference of impermeant solute molecules across a semi-permeable membrane produces a difference in solvent density, and in pressure, across the membrane. This has enormously important consequences - osmosis forms the basis of the transport of ions across cell membranes \cite{hille}, the regulation of blood salt levels by the kidneys \cite{naish}, uptake of water by plants \cite{raven, mauseth}, technologies for kidney dialysis  \cite{naish} and clean power production \cite{SteinErikSilhagen2007,boon}, and many other processes. Osmosis also lies at the  heart of many important phenomena in chemical physics including the Donnan effect \cite{biben_hansen}, the depletion interaction \cite{russel} and, recently, mechanisms for generating self-propelled colloidal particles \cite{Howe,Paxton1,Golestanian1,Golestanian2,Brady1,Kapral,Valeriani2010}. 

The cornerstone of our understanding of osmosis is the van 't Hoff relation, which states that a solute concentration difference, $\Delta c_{\mathrm{s}}$, leads to an equilibrium osmotic pressure difference $\Delta P = k_BT \Delta c_{\mathrm{s}}$ \cite{vantHoff1887,Huang}. Remarkably, this relation predicts that the osmotic pressure difference is  
the same as the pressure of an ideal gas at concentration $\Delta c_{\mathrm{s}}$, regardless of the molecular nature of the solute and solvent molecules. Gibbs showed that the van 't Hoff relation can be derived by setting the chemical potential of the solvent equal across the membrane \cite{Gibbs1897,Hill,Finkelstein,McMillanMeyer1945}; the thermodynamic origin of the osmotic pressure difference is then the entropy of mixing between solute and solvent. The van 't Hoff relation only holds for low solute concentrations; at higher concentrations, deviations from this ``ideal'' behaviour can be used as a diagnostic of solute-solute interactions, such as ion pairing \cite{atkins}. 

From a thermodynamic point of view, osmosis is thus well understood. However, osmotic effects are also important in non-equilibrium systems \cite{Howe,Paxton1,Golestanian1,Golestanian2,Brady1,Kapral,Valeriani2010,seminara,SteinErikSilhagen2007}, for which a thermodynamic description is not valid. For  systems that are out of equilibrium, a clear picture of the underlying molecular dynamical mechanisms is an essential prerequisite for understanding osmotic phenomena. Yet such a picture is largely lacking, even though the dynamical basis of osmosis has been the subject of over 100 years of discussion, since the seminal work of van 't Hoff. Relevant factors may include  bombardment of the membrane by solute molecules, diffusion of solvent across the membrane driven by its density difference, specific types of interaction between solute and solvent molecules and pulling of solvent molecules across the membrane in the wake of solute-membrane collisions \cite{SoodakIberall,OsmosisForum1,OsmosisForum2,OsmosisForum3,OsmosisForum4,OsmosisForum5,Dainty,nelson,ferrier,Yoffe}. 

While the overall picture remains unclear, progress has been made in understanding the dynamical basis of osmosis in specific systems. In an important contribution, Brady \cite{Brady} showed that for a system of colloidal solutes suspended in a coarse-grained solvent, the osmotic pressure can be derived from a purely mechanical, hydrodynamic Langevin description of colloidal motion; here, hydrodynamic interactions between colloids, mediated by solvent, play a central role. For systems with explicit solvent particles, Guell and Brenner \cite{Guell} showed that the van 't Hoff relation can be derived from first principles by taking into account all the forces involved; here, the role of membrane-solute interactions is emphasized. Importantly, while a thermodynamical description of osmosis cannot depend on the details of the system's dynamical rules (e.g. the interparticle interactions, membrane geometry, etc.), the nature of the underlying dynamical mechanisms may be sensitive to these details, and could be different for different osmotic systems. Indeed,  detailed models of osmotic water transport through membrane pores in ionic solutions have highlighted the importance of specific solvent-membrane and solvent-ion interactions in these systems \cite{RaghunathanAluru,RaghunathanAPL,Cannon}, while the  unusual characteristics of osmotic water flow through carbon nanotube membranes can be described by 1D stochastic hopping models \cite{Kalra2003}. Other studies have shown that partial solute penetration into membrane pores can strongly influence solvent flow \cite{Chou, Kim}, and that, for ionic solutes, non-trivial coupling can arise between the transmembrane electric field and osmotic flow \cite{aluru1,aluru2,Dzubiella}. 

In this paper, we seek to provide a clear ``minimal'' picture of the dynamical mechanisms underlying osmosis, in a system which is as simple as possible.  Murad and Powles \cite{Murad1993,Murad1994,Murad1995}, and Itano {\em{et al}} \cite{Itano}, have demonstrated that osmosis can be achieved in systems in which solute and solvent particles have identical Lennard-Jones interactions, while Luo and Roux \cite{luo} have used an external potential, invisible to the solvent, to mimic a semi-permeable membrane for ionic systems. Here, we combine these elements to create a ``minimal model'' osmotic system with purely repulsive particles, and with no solvent-membrane interactions, for which  we carry out a detailed study of the dynamics of solvent and solute particles in the osmotic steady state. For this system, we are able to derive a simple theoretical expression for the osmotic pressure in terms of solute-solute and solute-solvent interactions.  Analyzing the forces on the solvent particles in our system leads us to an intuitive model in which the solvent density and pressure gradients in osmotic equilibrium are maintained by a balance between an outward solvent flux generated by the osmotic pressure 
gradient and an inward diffusive flux. Although more complex mechanisms may be at play in real osmotic systems, they are not required to describe the fundamental physics of osmosis in this minimal model system. This work should provide a basis on which to build  future descriptions of osmotic phenomena in both equilibrium and non-equilibrium systems.

\section{A minimal model for osmosis}\label{sec:setup}

\begin{figure} 
 \includegraphics[width=0.4\textwidth]{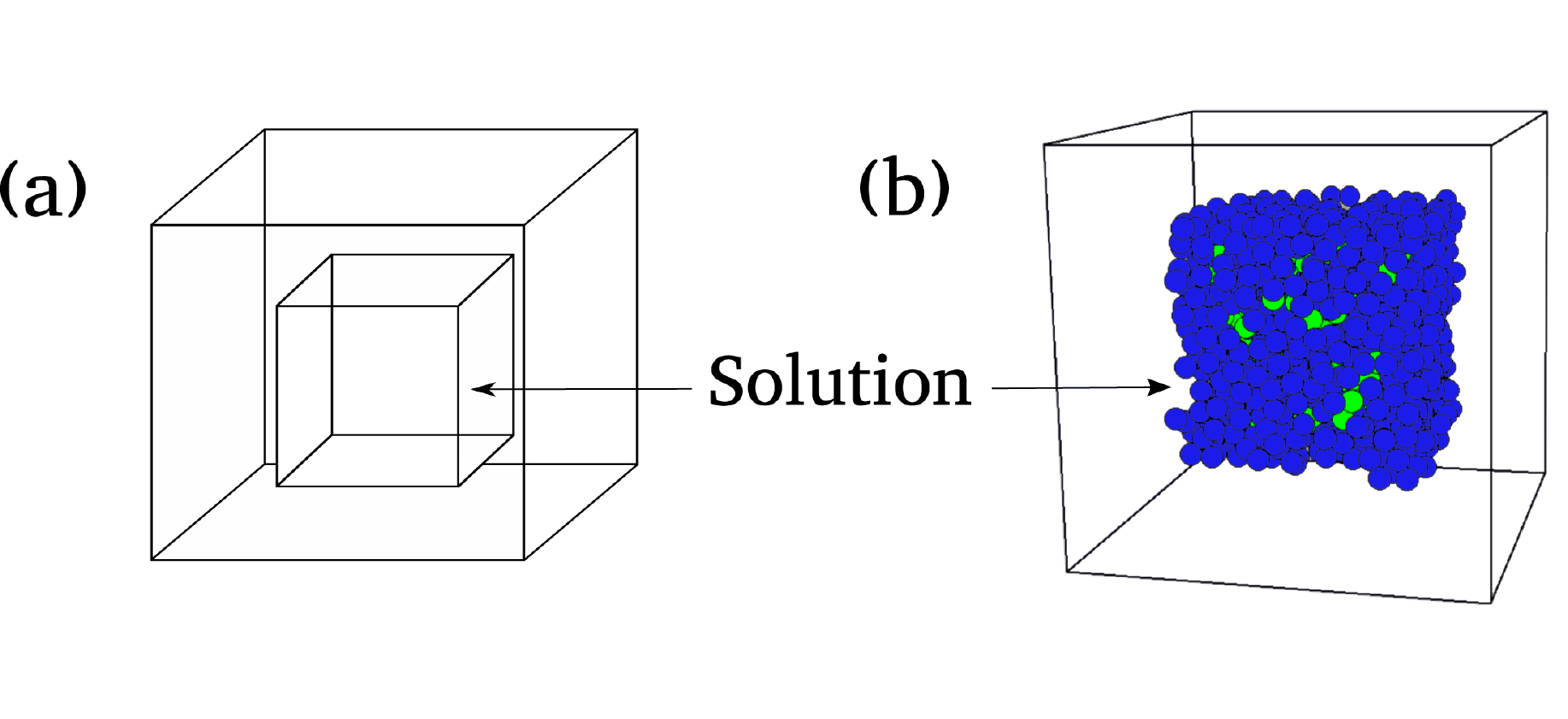}
 \caption{(a): Illustration of our model system. Solute particles are confined within a cubic solution compartment located in the centre of the simulation box. (b): Simulation snapshot; solute and solvent particles are coloured green and blue respectively. For clarity only particles located in the solution compartment are shown. }
 \label{fig:potential}
\end{figure}

Our model system is designed to reproduce the essential physics of osmosis, while eliminating as far as possible complications due to the specific chemical nature of the solutes, solvent and the semi-permeable membrane.  In our system solute and solvent particles interact with each other indistinguishably, and there are no solvent-membrane interactions. Our simulation setup is illustrated in Figure \ref{fig:potential}. The simulation box is partitioned into ``solution'' and ``solvent'' compartments; the solute particles are confined within the central solution compartment by an external potential of the form $U = k_BT(\sigma/dx)^{9}$ (where $dx$ is the perpendicular distance between a solute particle and the boundary of the compartment, and $\sigma$ is the particle diameter) \footnote{For the corners, the interactions with each wall are evaluated separately and summed.}. The solvent particles do not experience the confining potential and are free to move throughout the  simulation box. The confining potential therefore acts like a semi-permeable membrane. Both solvent and solute particles are of unit mass $m$ and interact via identical, repulsive,  Weeks, Chandler, Andersen (WCA) interactions ($U(r) = 4\epsilon \left[\left(\frac{\sigma}{r}\right)^{12}-\left(\frac{\sigma}{r}\right)^{6}+\frac{1}{4}\right]$ if $r<2^{1/6}\sigma$ and zero otherwise), with parameters $\sigma = 1$ and $\epsilon = k_BT = 1$ \cite{WCA1971} (i.e. our units of energy and distance are $k_BT$ and $\sigma$ respectively). The system is simulated using Molecular Dynamics, with the velocity Verlet algorithm \cite{AllenTildesley,FrenkelSmit} and timestep 0.001 (in reduced units \cite{AllenTildesley}, \footnote{The reduced unit of time is ($m\sigma^2/\epsilon)^{1/2}$; in our simulations $m$, $\sigma$ and $\epsilon$ are all set to unity.}), combined with a Nos{\'{e}}-Hoover thermostat \cite{Nose1984,AllenTildesley}. In total, our system contains  5000 particles (solute plus solvent) in a periodic, cubic simulation box of size $L=18.42\sigma$, so that the total particle density $\rho_{\mathrm{total}}=0.8 \sigma^{-3}$ (the packing fraction $\pi \rho_{\mathrm{total}}/6 = 0.42$). 

In this paper, we focus on the osmotic steady state; we therefore allow the system to equilibrate thoroughly  before data is collected \footnote{The equilibration time and run time are at least 40 and 2000 reduced time units respectively.}. We measure the local pressure in the solution and solvent compartments using the  Method of Planes \cite{IrvKirkPres,ToddEvansMOP}. As discussed in Appendix \ref{app:MOP}, this constitutes a direct measurement of the kinetic and interaction components of the momentum flux across a local plane. The osmotic pressure is then computed by taking the difference between the pressures in the two compartments. We could also have measured the osmotic pressure directly from the force exerted by the solute particles on the membrane (as done by Luo and Roux \cite{luo}), or using virial expressions for the local pressure \cite{Lion2012}; these methods give identical results. To compute the concentration of solute particles, $c_{\mathrm{s}}$, we need to  define the volume of the solution compartment. This is non-trivial, because the confining potential is smooth. We define the solution volume by matching the pressure-density relation for a gas of solute particles, confined in the solution compartment, in the absence of solvent, with that of a system of WCA particles simulated in a periodic box (see Appendix \ref{app:MOP}). 

\section{Osmosis in the model system}

\subsection{Density imbalance}

\begin{figure*}[!t]
\includegraphics[width=0.65\textwidth]{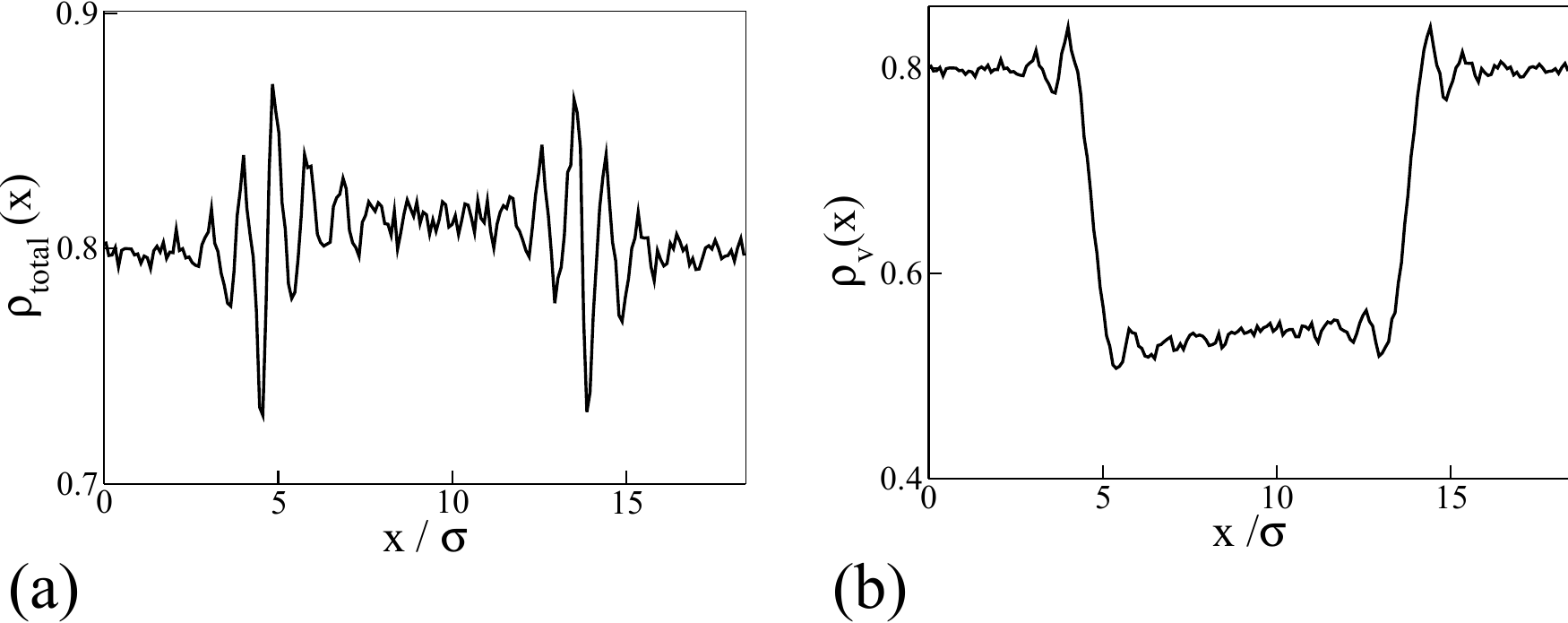}
\caption{Local density profiles $\rho(x)$ (in units of $\sigma^{-3}$) measured across the middle of the simulation box, for solute concentration $c_{\mathrm{s}} = 0.254\sigma^{-3}$. Panel (a) shows the 
total particle density, $\rho_{\mathrm{total}}(x)$; panel (b) shows the local solvent density $\rho_{\mathrm{v}}(x)$. }
 \label{fig:density1}
\end{figure*}

\begin{figure}[!b]
\makebox[10pt][l]{} \includegraphics[width=0.4\textwidth]{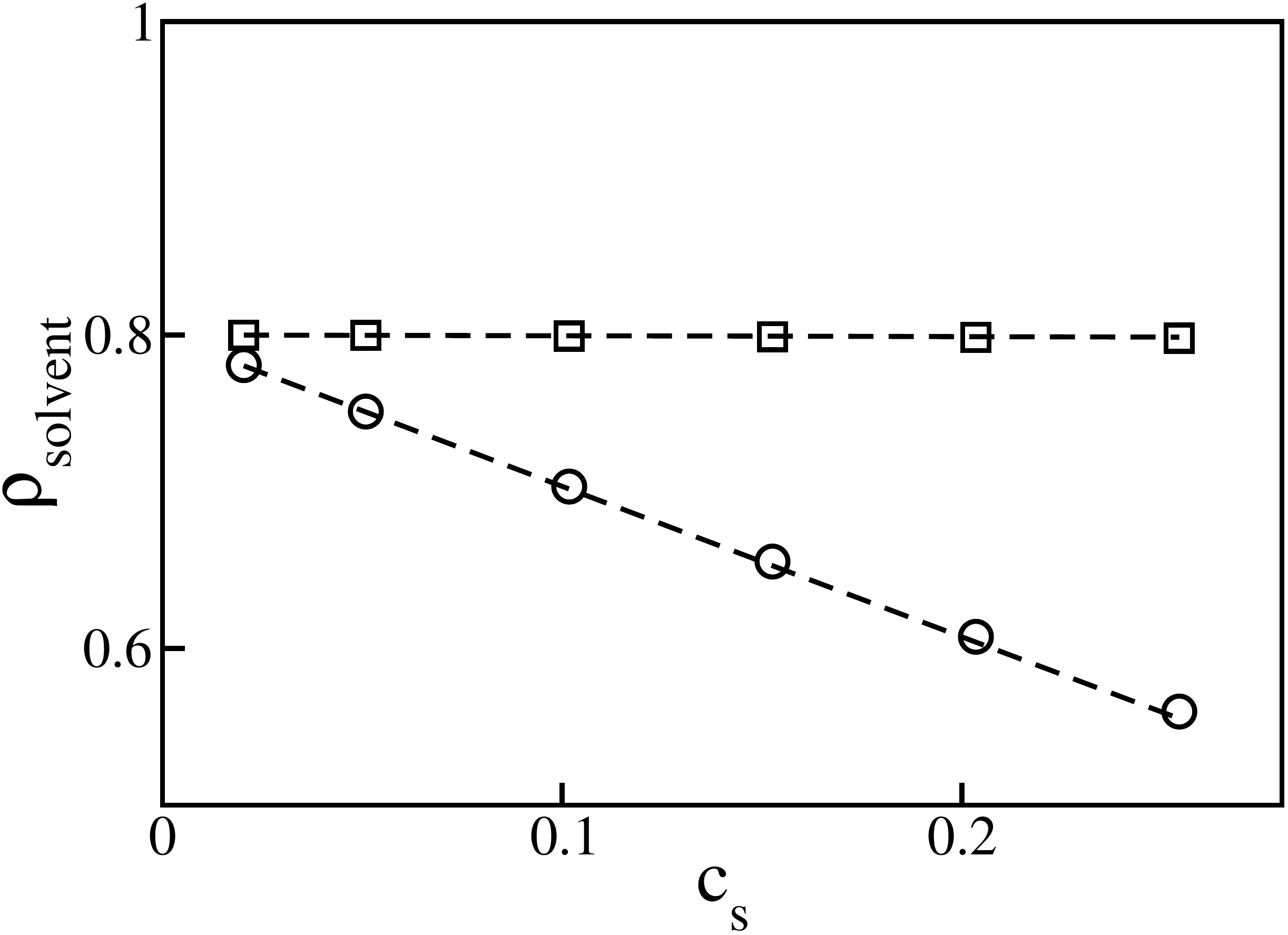}
 \caption{Spatially averaged solvent density $\rho_{\mathrm{v}} $ as a function of solute concentration $c_{\mathrm{s}}$ (both in units of $\sigma^{-3}$), in the solution and solvent compartments (circles and squares respectively). The dashed lines show theoretical predictions based on the Carnahan-Starling equation of state (see Appendix \ref{app:CS}).}
 \label{fig:density2}
\end{figure}

We first verify that osmosis indeed occurs in our model system by measuring the steady-state local density profiles in the solution and solvent compartments. Figure \ref{fig:density1} shows density profiles, taken through the middle of our simulation box. As expected, the {\em{total}} particle density, solute plus solvent (Figure \ref{fig:density1}a), 
is higher in the solution than in the solvent compartment. In contrast, the {\em{solvent}} density is lower in the solution than in the solvent compartment (Figure \ref{fig:density1}b). This density imbalance increases linearly with the concentration of solute, $c_{\mathrm{s}}$, as shown in Figure \ref{fig:density2}. These results are in good agreement with thermodynamic predictions, obtained by approximating the WCA particles as hard spheres and setting the solvent chemical potential, obtained from the Carnahan-Starling equation of state, equal in the two compartments (dashed lines in 
Figure \ref{fig:density2}; for details of the calculations see Appendix \ref{app:CS}). Figure \ref{fig:density2} also demonstrates that  over the parameter range considered here, the particle density in the solvent compartment remains virtually unaffected by changes in the solute concentration, confirming that the solvent compartment is large enough to be regarded as a reservoir. 

\subsection{Osmotic pressure}
\label{sec:OPresults}	

\begin{figure}[!hb]
 \includegraphics[width=0.4\textwidth]{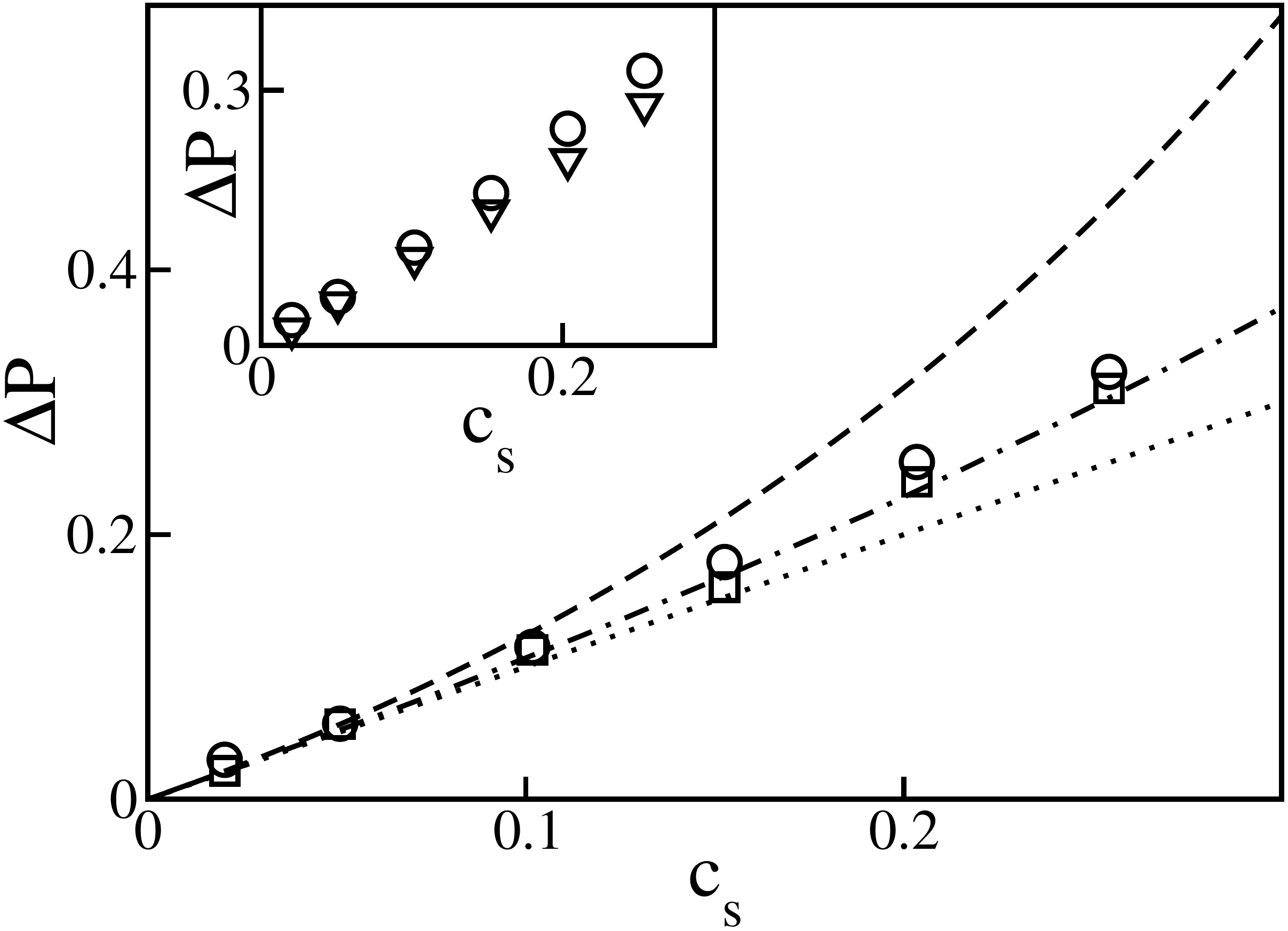}
 \caption{Osmotic pressure $\Delta P$ (in units of $k_BT\sigma^{-3}$) as a function of solute concentration $c_{\mathrm{s}}$ (in units of $\sigma^{-3}$). In the main plot, the symbols show simulation results. The circles show direct measurements of $\Delta P$ in our simulations, computed using the method of planes (see Appendix \ref{app:MOP}). The squares show  $\Delta P$ computed from our simulations using Eq.~(\ref{eq:vHpluscorr}). The statistical errors on the symbols are $\pm 3\%$ for $c_s$ (circles and squares) and $\pm 3\%$ for $\Delta P$ (squares only) -- i.e. approximately the size of the symbols. These errors arise mainly from uncertainty in the position of the solution boundary, as discussed in Appendix  \ref{app:MOP}. The lines show theoretical predictions. The dotted line shows the van 't Hoff relation, $\Delta P = k_BT c_{\mathrm{s}}$. The dashed line shows the pressure of a system of hard spheres at density $c_{\mathrm{s}}$, computed using the Carnahan-Starling equation of state. The dot-dashed line shows the osmotic pressure predicted by a full thermodynamic calculation, including solute and solvent, using the  Carnahan-Starling equation of state. In the inset, the circles are the same as in the main plot, while the triangles show the prediction of our simple hopping model (Eq.(\ref{eq:hopping1})) $\Delta P = -k_BT \rho_{\mathrm{v}}^{\mathrm{out}}\log{\left[\rho_{\mathrm{v}}^{\mathrm{in}}/\rho_{\mathrm{v}}^{\mathrm{out}}\right]}$ (where $\rho_{\mathrm{v}}^{\mathrm{in}}$ and $\rho_{\mathrm{v}}^{\mathrm{out}}$ are the average solvent densities in the solution and solvent compartments, respectively). 
}
 \label{fig:dP-vs-conc}
\end{figure}

The osmotic pressure, $\Delta P$ (i.e. the pressure difference between the solution and solvent compartments), is shown in Figure \ref{fig:dP-vs-conc} as a function of the solute concentration, $c_{\mathrm{s}}$. At low solute concentration, our simulation results (circles) are in good agreement with the  van 't Hoff relation (dotted line). At high solute concentration ($c_{\mathrm{s}} > 0.1 \sigma^{-3}$, corresponding to solute packing fraction greater than $0.05$), the osmotic pressure exceeds that predicted by the van 't Hoff relation, as expected in a system with repulsive solute-solute and solute-solvent interactions. Interestingly, however, the osmotic pressure in our system is significantly lower than the pressure that would be obtained for a gas of WCA particles at density $c_{\mathrm{s}}$ (dashed line in Figure \ref{fig:dP-vs-conc}; computed using the Carnahan-Starling equation of state). Thus a na{\"{i}}ve picture in which one treats the solution simply as a ``solute gas'', ignoring the solvent,  does not correctly account for the osmotic pressure. Figure \ref{fig:dP-vs-conc} also shows the osmotic pressure predicted by a full thermodynamic calculation, for solvent and solute, using the Carnahan-Starling equation of state (dot-dashed line); this prediction agrees well with our simulations, as expected for a system of particles with hard sphere-like interactions. 

Standard derivations of the osmotic pressure (including the dot-dashed line in Figure \ref{fig:dP-vs-conc}) rely on thermodynamic principles -- i.e. on equality of the chemical potential across the membrane. In contrast, our aim in this paper is to investigate the {\em{mechanical}} origins of osmosis. It is thus of interest to derive expressions for the osmotic pressure purely from mechanical principles, without recourse to thermodynamics. In Appendix \ref{app:virial} we present one such derivation. Starting from the Clausius virial relation for the solute particles, we derive a (to our knowledge new) approximate expression for the osmotic pressure in terms of the forces of interaction between the solute and solvent particles: 

\begin{equation}
 \Delta P \approx k_{B}T c_{\mathrm{s}} + \frac{1}{3V}\left(\sum_{\mathrm{i_s}} \sum_{\mathrm{j_s>i_s}}\vec r_{i}.\vec f_{ij} + \sum_{\mathrm{i_s}}\sum_{\mathrm{j_v}} \vec r_{i}.\vec f_{ij} \right).
 \label{eq:vHpluscorr}
\end{equation}

Here, $V$ is the volume of the solution compartment (defined as described above), $\vec r_{i}$ is the position of particle $i$ and $\vec f_{ij}$ is the force exerted on particle $i$ by particle $j$. The first term in the brackets sums over all pairs of solute particles (avoiding double counting) while the second term sums over all solute-solvent pairs (note that here particle $i$ denotes the solute and $j$ the solvent; the solvent may be inside or outside the solution compartment). Since the starting point of this derivation, the Clausius virial relation, follows from Newton's equations of motion in a system at steady state (see Appendix \ref{app:virial}), Eq.~(\ref{eq:vHpluscorr}) amounts to a purely mechanical description of the osmotic pressure in our system. Figure \ref{fig:dP-vs-conc} (squares) shows that Eq.~(\ref{eq:vHpluscorr}) is indeed in good agreement with the direct measurement of $\Delta P$ from our simulations (circles), over the full range of solute concentrations tested. 

Expression (\ref{eq:vHpluscorr}) makes an interesting connection with ``effective single-component'' descriptions of osmotic systems, as used, for example, for colloidal dispersions. Here, one aims to coarse-grain the system of solute and solvent particles, representing it by a  single-component fluid of solute particles, which interact via an effective potential that takes into account the effects of the solvent. This works well for colloidal dispersions where the colloids (solutes) are  orders of magnitude larger than the solvent molecules and the effective interactions are well-represented by a pairwise intercolloidal potential $V(r)$. Here, by analogy with atomic systems the osmotic pressure, $\Pi$, is written as $\Pi = nk_BT - \frac{2\pi}{3}n^2\int_0^\infty r^3 g(r) \left(dV/dr\right) dr$, where $n$ is the number density of colloidal particles and $g(r)$ is the radial pair distribution function \cite{russel,JPHansen,Brady}. This expression can also be derived from a purely  dynamical description of colloidal motion, taking account of Brownian motion and hydrodynamic interactions \cite{Brady}. Eq.~(\ref{eq:vHpluscorr}) provides an analogous expression for the osmotic pressure, for a system in which the solvent degrees of freedom are treated explicitly and on the same footing as those of the solute.

\section{What maintains the solvent density gradient?}

\subsection{Balance between outward and inward fluxes}
\label{sec:rhograd}

\begin{figure}[!b]
\includegraphics[width=0.4\textwidth]{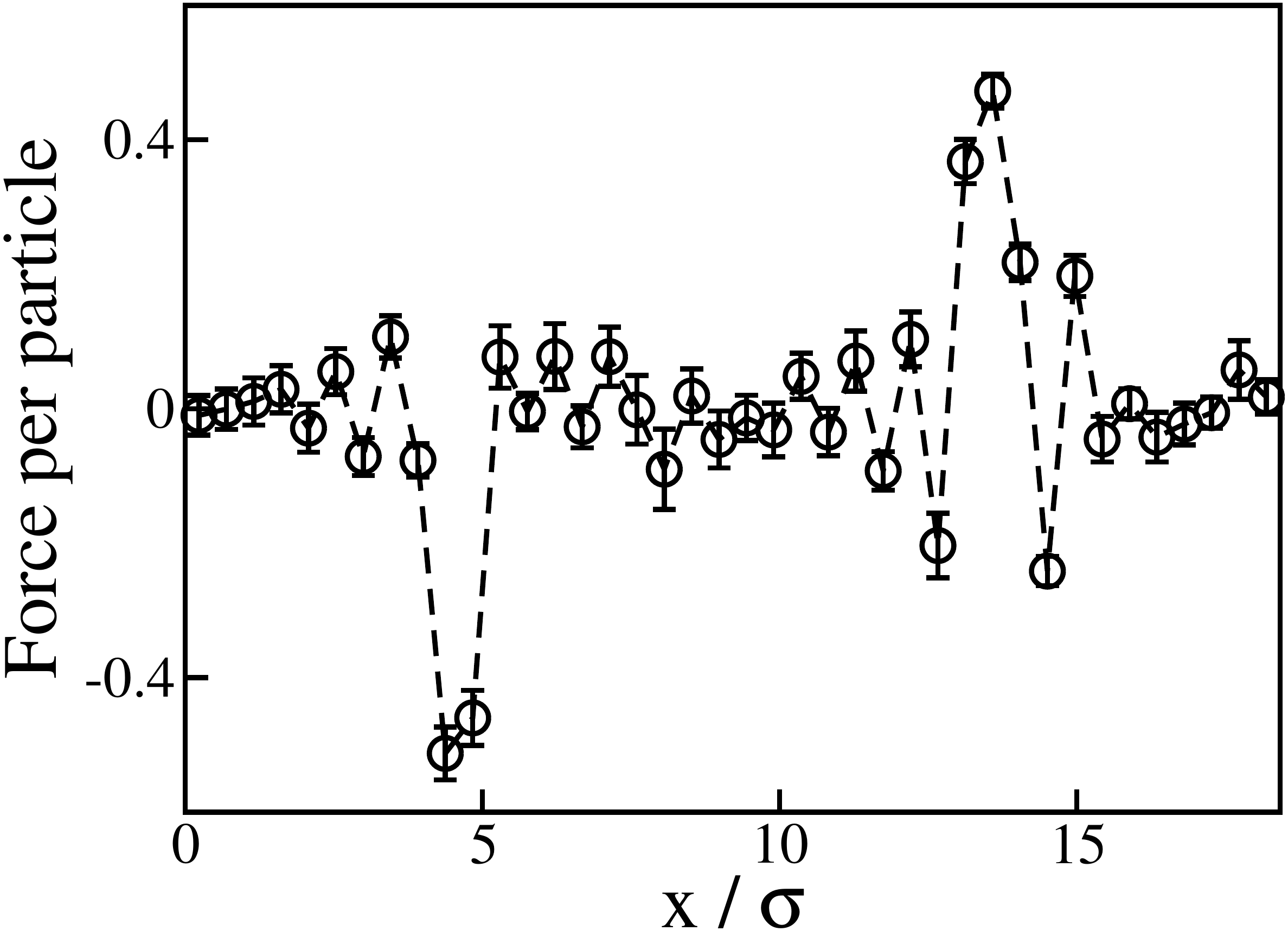}
\caption{Net force per particle, normal to the membrane (in units of $k_BT\sigma^{-1}$) acting on the solvent particles, as a function of position $x$, in a slab taken through the middle of the simulation box, for solute concentration $c_{{\mathrm{s}}} = 0.254 \sigma^{-3}$. Solvent particles close to the boundaries of the solution compartment tend to be pushed out of the solution. }
\label{fig:force}
\end{figure}

Our simulations allow us to investigate in detail the microscopic forces which produce the imbalance in the density of solvent particles between the solution and solvent compartments. Figure \ref{fig:force} shows the net force per particle, acting on the solvent particles, as a function of position $x$, in a slab through the middle of the simulation box. Solvent particles close to the boundaries of the solution compartment are on average pushed out of the solution, towards the solvent compartment. This net force arises from the fact that the total density inside the solution compartment is higher than that in the solvent compartment (as shown in Figure \ref{fig:density1}a) -- thus, we expect solvent particles at the boundary to experience more collisions from the solution side than from the solvent side. 

Why then is there no net flow of solvent out of the solution compartment in response to the net outward force? The answer lies in the solvent density profiles. As shown in Figure \ref{fig:density1}b, the density of {\em{solvent}} particles is higher in the solvent compartment than in the solution. We expect this to create a diffusive flux of solvent particles  into the solution, which counteracts the outward flux caused by the net outward force. Considering, rather simplistically, the ``hopping'' of solvent particles between layers next to the membrane on the solvent and solution sides, we expect that an 
 individual particle on the solution side has a high probability of leaving the solution (i.e. to hop from the solution to the solvent compartment), due to the outward pushing force, while an individual particle on the solvent side has only a low probability of entering the solution (i.e. to hop from the solvent to the solution compartment). However, there are more 
solvent particles on the solvent side and consequently more attempts to hop into the solution than out of it. Thus the net fluxes of solvent particles inward and outward are equal and the solvent density imbalance is maintained. This picture has much in common with the classic colloidal sedimentation equilibrium, in which an inhomogeneous colloidal density profile is maintained by a balance of a downward flux driven by the gravitational field and an upward diffusive flux. This analogy between osmosis and sedimentation was also noted by Soodak and Iberall \cite{SoodakIberall}. We note that these density profiles are also consistent with the statistical mechanical relation between the density and the potential of mean force \cite{Hill, JPHansen}. However, the latter relation does not contain any dynamical information.

Using this simple ``hopping'' picture, we can  relate the solvent density imbalance to the osmotic pressure by simple mechanical arguments. Let us assume the solvent densities, $\rho_{\mathrm{v}}^{\mathrm{in}}$ and $\rho_{\mathrm{v}}^{\mathrm{out}}$ in the solution and solvent compartments respectively, are uniform, with a very sharp density change at the boundary. We now define slabs of width $h$ (of the order of the molecular size) next to the boundary on the solvent and solution sides, and consider the hopping of solvent particles between these slabs. Solvent particles that hop from the solvent side into the solution experience an energy penalty $h \vec{f}_{\mathrm{out}}$, where the outward force $\vec{f}_{\mathrm{out}}$ arises from the fact that the total density is higher in the solution, and can be related to the osmotic pressure $\Delta P$ by $\Delta P = \vec{f}_{\mathrm{out}}/a$, where $a$ is the slab area per  particle. Setting the fluxes of solvent particle hops equal in the inward and outward directions, we find that $\rho_{\mathrm{v}}^{\mathrm{in}} = \rho_{\mathrm{v}}^{\mathrm{out}}\exp{\left[-\vec{f}_{\mathrm{out}}h/(k_BT)\right]}= \rho_{\mathrm{v}}^{\mathrm{out}}\exp{\left[-ah \Delta P /(k_BT)\right]}$. Noting now that $ah$ is the volume per particle, which is approximately the same on both sides of the membrane -- i.e. that $ah \approx 1/ \rho_{\mathrm{total}}^{\mathrm{in}} \approx 1/\rho_{\mathrm{v}}^{\mathrm{out}}$, leads to the result

\begin{equation}
\Delta P \approx -k_BT  \rho_{\mathrm{v}}^{\mathrm{out}} \log{\left[\frac{\rho_{\mathrm{v}}^{\mathrm{in}}}{\rho_{\mathrm{v}}^{\mathrm{out}}}\right]} .
\label{eq:hopping1}
\end{equation}

Eq.~(\ref{eq:hopping1}) is in fact a standard relation, which can be derived via thermodynamic arguments \cite{Finkelstein,smith}, by setting the solvent chemical potential equal across the membrane and making the assumption that $\rho_{\mathrm{total}}^{\mathrm{in}} \approx \rho_{\mathrm{v}}^{\mathrm{out}}$. Thus our simple hopping model provides a clear mechanistic description which leads to the same predictions as the thermodynamic/statistical mechanical pictures. The mechanistic description proposed here of course relies on the presence of repulsive interparticle interactions, which generate the ``outward pushing force''. In a hypothetical case where the solute and solvent particles do not interact with each other (``ideal gas''), this argument would not hold. However in this case there would also be no osmotic 
density imbalance; the solvent density would be equal on both sides of the membrane \cite{kiil}. 

Figure \ref{fig:dP-vs-conc} (inset; triangles) shows that Eq.~(\ref{eq:hopping1}) holds in our simulations, even for solute concentrations well beyond the limit of validity of the van 't Hoff relation. The slight discrepancy between the prediction of Eq.~(\ref{eq:hopping1}) and the simulation results arises from the assumption $\rho_v^{out} \approx \rho_{total}^{in}$ made in our derivation. 

For low solute concentrations, Eq.(\ref{eq:hopping1}) reduces to the van 't Hoff relation. Noting that $\rho_{\mathrm{v}}^{\mathrm{in}} = \rho_{\mathrm{total}}^{\mathrm{in}} - c_{\mathrm{s}} \approx \rho_{\mathrm{v}}^{\mathrm{out}} - c_{\mathrm{s}}$ leads to: 

\begin{equation}
\log{\left[1-\frac{c_{\mathrm{s}}}{\rho_{\mathrm{v}}^{\mathrm{out}}}\right]} \approx -\frac{\Delta P }{k_BT \rho_{\mathrm{v}}^{\mathrm{out}}}.
\label{eq:hopping2}
\end{equation}

The van 't Hoff relation is then recovered when we expand  the logarithm to first order in $c_{\mathrm{s}}/\rho_{\mathrm{v}}^{\mathrm{out}}$ (which is small for low solute concentration):

\begin{equation}
\Delta P \approx -k_BT \rho_{\mathrm{v}}^{\mathrm{out}} \log{\left[1-\frac{c_{\mathrm{s}}}{\rho_{\mathrm{v}}^{\mathrm{out}}}\right]} \approx k_BT c_{\mathrm{s}}.
\label{eq:hopping3}
\end{equation}

\subsection{Solvent-solute correlations}

\begin{figure*}
 \includegraphics[width=0.65\textwidth]{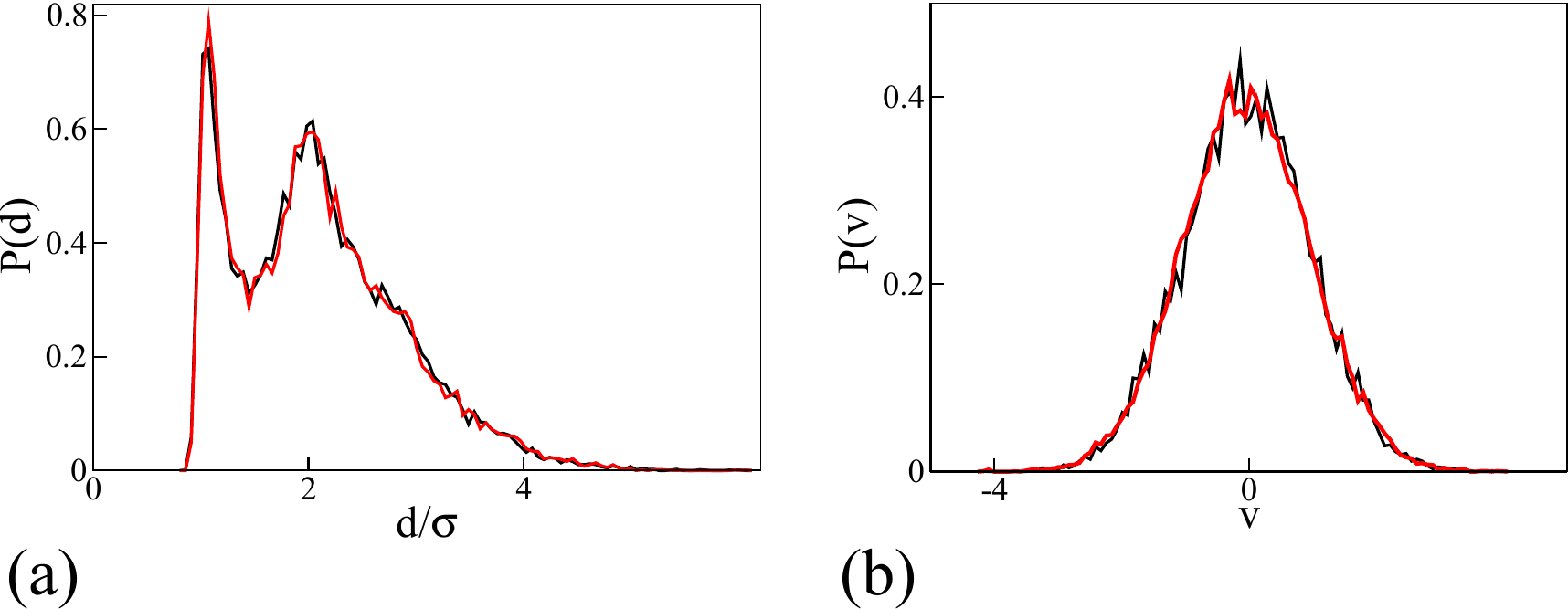}
 \caption{Probability distributions for the distance $d$ to the nearest solute particle (a) and the perpendicular  velocity $v$ of the nearest solute particle (b), for solvent particles located less than 0.001 particle diameters outside the solution compartment, which do (black lines) or do not (red lines) enter the solution compartment in the next simulation timestep. In these simulations, $c_{\mathrm{s}}=0.05 \sigma^{-3}$. The boundary of the solution compartment is defined as described in Appendix \ref{app:MOP}.}
 \label{fig:velocities}
\end{figure*}

The simple picture presented above provides a mechanical description which is consistent with results obtained from thermodynamic arguments, and which is also consistent with the force profiles measured in our simulations. To show that this  actually describes what is happening in our simulations, we need to rule out  more complex dynamical mechanisms. A key feature of our simple picture is that it does not predict any correlations between the configuration of the solute particles and the hops of solvent particles across the membrane - unlike some other proposed mechanisms in which solvent particles are pulled across the membrane in the wake of nearby solute particles. To look for such effects in our simulations, we therefore test whether  correlations between solute and solvent motion influence the dynamics of solvent particles close to the membrane. 

We first ask whether the microscopic flux of solvent particles into the solution compartment is influenced by the presence of nearby solute particles. At time $t$, we focus on solvent particles which are situated less than 0.001 particle diameters\footnote{This corresponds to a similar distance to  that which a particle at temperature $T = 1$ will move in one timestep.} outside the solution compartment. Within the next timestep, some of these solvent particles cross into the solution compartment, while others do not. Figure \ref{fig:velocities}a shows the distribution of distances to the nearest solute particle, for those solvent particles which do cross in the next timestep (black line) and for those which do not (red line). These distributions are indistinguishable, suggesting that proximity of a solute particle does not make any difference to the chance that a solvent particle, located close to the boundary, will cross into the solution compartment. 

One might also suppose that the {\em{velocity}} of nearby solute molecules might influence the microscopic flux of solvent particles -- e.g. a solvent particle could cross into the solution in the wake of  a nearby solute particle as it rebounds from the membrane. Figure \ref{fig:velocities}b shows the probability distribution for the velocity of the nearest solute particle (perpendicular to the solution boundary), for solvent particles located close to the solution boundary, which do and do not subsequently cross into the solution. Once again, the two probability distributions are indistinguishable.

\begin{figure}[!b]
\includegraphics[width=0.4\textwidth]{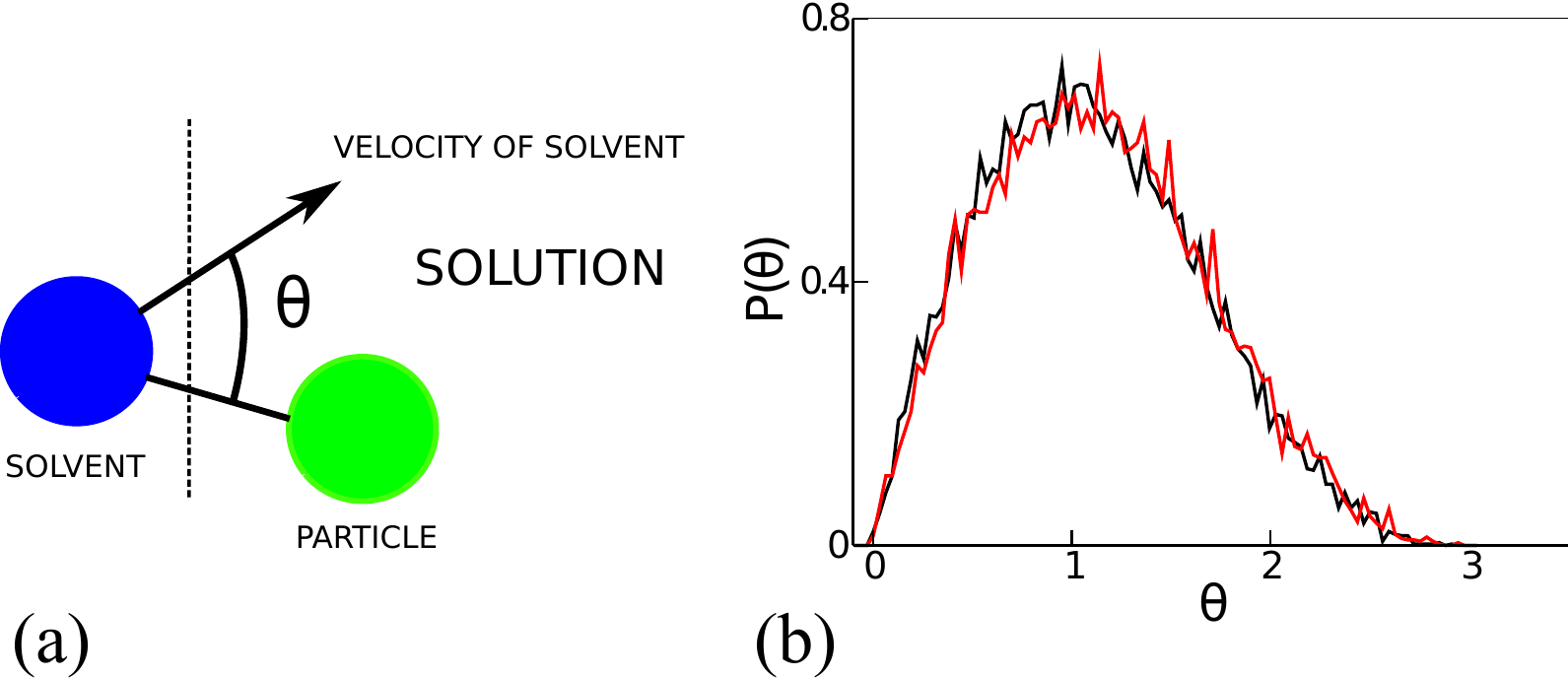}
 \caption{(a): $\theta$ is defined as the angle between the instantaneous velocity of a solvent particle $i$ as it enters the solution compartment and the vector joining it to particle $j$ within the solution compartment. (b): Probability distribution $P(\theta)$ when particle $j$ is defined to be the nearest solute particle (black line) and when particle $j$ is the nearest particle of either solute or solvent (red line). In these simulations  $c_{s} = 0.05\sigma^{-3}$.  The boundary of the solution compartment is defined as described in Appendix \ref{app:MOP}. }
 \label{fig:angles}
\end{figure}

As a final test, we measured whether solvent crossings occur preferentially in the {\em{direction}} of the nearest solute particle. If solute particles tend to ``suck'' solvent particles across the membrane, solvent crossings should be preferentially oriented towards the nearest solute particle. We therefore measured the angle $\theta$ between the instantaneous velocity $\vec{v}_i$, of a solvent particle as it crossed into the solution compartment, and the vector $\vec{r}_{ji} = \vec{r}_{j}-\vec{r}_{i}$, where $\vec{r}_{i}$  and $\vec{r}_{j}$ are the position of the solvent particle and the nearest solute, respectively (see Figure \ref{fig:angles}a). Figure \ref{fig:angles}b (black line) shows the probability distribution $P(\theta)$. We then repeated the calculation, but this time allowing particle $j$ to be either solute or solvent (within the solution compartment). The resulting distribution is given by the red line in Figure \ref{fig:angles}b. Again, the two distributions are indistinguishable, showing that solvent particles do not cross the membrane preferentially in the direction of nearby solute particles. We therefore conclude that  correlations between solute and solvent dynamics do not play a significant role in determining the microscopic  solvent flux in our simulations. This suggests that the simple dynamical picture described above is indeed sufficient to describe osmosis in our system.

\section{Discussion}

In this paper, we have used a minimal model to investigate the dynamic mechanisms underlying the maintenance of equilibrium osmotic pressure and density gradients. In our system, solute and solvent particles interact identically, and there are no solvent-membrane interactions. For this system, we are able to derive an expression (Eq.(\ref{eq:vHpluscorr})) for the osmotic pressure, in terms of the forces of interaction between solute-solute and solute-solvent pairs of particles; this derivation does not assume any separation of scales between the solvent and solute particles \footnote{Of course, one is always free to coarse-grain the solvent particles, treating the solutes as a ``gas'' with effective interactions which account for the presence of the solvent -- but for systems like that treated here without solvent-solute scale separation, these effective interactions may be many-body}. 

Analysis of the density and force profiles in our simulations leads us to propose a very simple picture, in which the density imbalance of solvent particles across the membrane is maintained by a balance between an outward force-driven flux (due to the  higher total density in the solution) and an inward diffusive flux (due to the lower solvent density in the solution). We show that a simple calculation, based on the mechanics of hopping of solvent molecules across the membrane, can be used to derive the same relation  between the solvent density gradient and the osmotic pressure (Eq.(\ref{eq:hopping1})) as is obtained by standard thermodynamic arguments. For low solute concentrations, this expression reduces to the van 't Hoff relation. We do not detect any evidence of correlations between the solute particles and the solvent flux, leading us to believe that this simple ``hopping'' picture is sufficient to describe the dynamical basis of osmosis in our simulations. 

In previous work, various molecular mechanisms have been implicated in osmosis, including solute-membrane collisions, diffusion of solvent across the membrane, solute-solvent interactions and correlations between solvent and solute dynamics. In simulations of realistic osmotic systems, the dynamics of solvent molecules inside pores \cite{Kalra2003,Allen2003}, interactions with the pore wall \cite{RaghunathanAluru}, interactions between solvent and solute molecules \cite{RaghunathanAluru,Cannon}, and correlations between  solvent flux and solute dynamics \cite{Chou} have all been shown to play important roles. It is possible that different dynamical mechanisms are at play in different osmotic systems, all of these being consistent with the known, and universal, thermodynamic relations. Thus, an important topic for further work will be to understand how the conclusions of our work change for more realistic scenarios: e.g. when the membrane presents a barrier also to the solvent molecules (via explicit pores or via a smooth potential barrier), when the solute and solvent do not interact identically, etc. Our aim in this work, however, was to focus on a minimal model system, for which we could characterize in detail the underlying molecular mechanisms. For this system, we find that complex mechanisms are not required to understand the basic physics of osmosis.

We hope that this study will provide a basis for developing our understanding of osmosis in systems out of equilibrium, for which thermodynamic concepts are not available. Recently, much attention has focused on active particles such as  
motile bacteria, swimming ``Janus'' colloids or colloidal ``chuckers'' \cite{Howe,Paxton1,Golestanian1,Golestanian2,Brady1,Kapral,Valeriani2010}. These can act as solutes: an interesting question concerns how the osmotic pressure of a suspension of active colloids differs from that of the corresponding passive suspension. Moreover, active particles may themselves generate osmotic gradients, with interesting potential for motility and self-organisation. The approach described here can easily be extended to investigate the dynamics of osmotic systems in which either the solute or the solvent particles are active. 

Finally, we note that the minimal setup presented in this paper also provides a very general way to measure osmotic pressure in systems involving solvent and solutes, simply by partitioning a system into two compartments using an external potential permeable only to the solvent. Indeed, a similar approach has already been used to fit intermolecular potentials to experimental data on osmotic pressure \cite{luo}. This kind of setup should facilitate osmotic pressure measurements in instrinsically non-equilibrium systems (e.g. with self-propelled particles), where thermodynamic approaches are not feasible. 

\begin{acknowledgments}

The authors thank Daan Frenkel for drawing this topic to our attention, Leo Lue for assistance with the calculations in Appendix \ref{app:CS},  Mike Cates,  Davide Marenduzzo and Aron Yoffe for useful discussions and Mike Cates and Patrick Warren for helpful comments on the manuscript. TWL was supported by an EPSRC DTA studentship and RJA by a Royal Society University Research Fellowship and by EPSRC under grant EP/EO30173. This work has made use of the  Edinburgh Compute and Data Facility, which is partially supported by the eDIKT initiative.

\end{acknowledgments}

\begin{appendix}

\section{Pressure measurement and solution boundary}\label{app:MOP}

We measure the local pressure in the solution and solvent compartments using the method of planes \cite{IrvKirkPres,ToddEvansMOP}. In this method, one  defines a plane within the region of interest, and monitors the  flux $\phi$ of particle momentum normal to the plane (due to particles crossing the plane), and the interparticle forces normal to the plane, for pairs of particles on opposite sides of the plane. For a plane at position $x=x'$, which extends right across the 
simulation box in the $y$ and $z$ directions, the pressure $P$ is given by
 
\begin{equation}
 P = \phi + \frac{1}{2A}\sum_{i,j > i}[sgn(x_{i}-x')-sgn(x_{j}-x')] f_{xij},
 \label{eq:MOP}
\end{equation}

where $A$ is the area of the plane, the sum is over all pairs of particles (without double counting),  $x_i$ and  $x_j$ are the x-components of the particle position (${\bf{\hat{x}}}$ being the normal to the plane) and $f_{xij}$ is the normal component of the force between particles $i$ and $j$. The term in the square brackets ensures that only pairs of particles which are on opposite sides of the plane contribute to  $P$. Expression (\ref{eq:MOP}) describes the instantaneous pressure, which has $\delta$-function peaks whenever a particle crosses the plane; this expression is time-averaged to provide a meaningful pressure measurement. 

\begin{figure}[t]
 \includegraphics[width=0.4\textwidth]{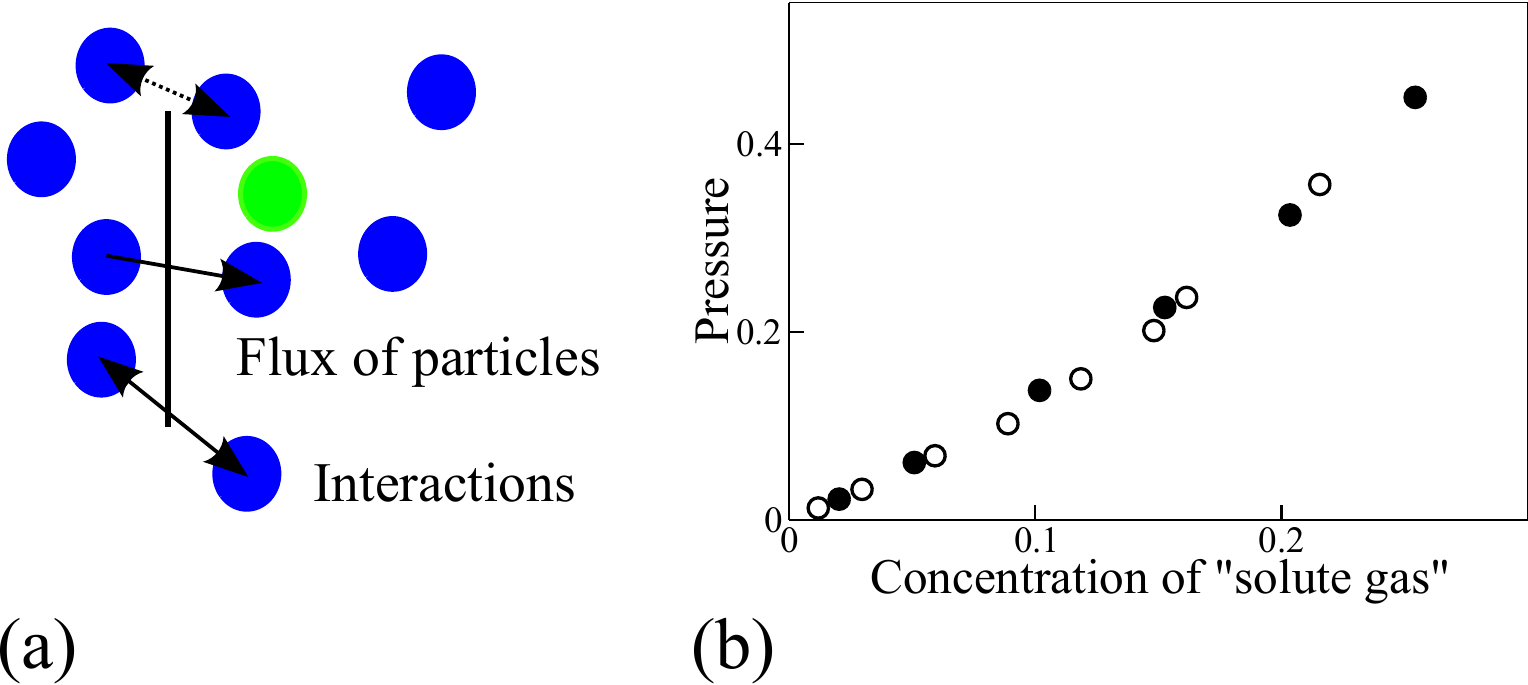}
 \caption{(a): Method of planes with a plane of finite extent in the $y$ and $z$ directions. Solid unidirectional arrows denote contributions to the kinetic part of Eq.(\ref{eq:MOP}) ($\phi$); solid bidirectional arrows denote contributions to the interaction part of Eq.(\ref{eq:MOP}). All interactions in which the line connecting the two particles crosses the plane are included (e.g. the bottom pair of particles); however the interaction between the top pair of particles (dotted arrow) is excluded. (b): Pressure-density relation of a ``gas'' of solute particles, confined in the solution compartment in the absence of solvent (closed circles), compared with that of a periodic box of equivalent particles (open circles). The pressure is in units of $k_BT \sigma^{-3}$; the density is in units of $\sigma^{-3}$. For the closed circles, the density of the solute gas is defined using an effective boundary of the solution compartment positioned $0.91\sigma$ inside the locus of divergence of the the confining potential.} 
\label{fig:MOPdiag}
\end{figure}

In our simulations, we modify this method in order to use planes with finite extent in the $y$ and $z$ directions. In this case, the sum in Equation \ref{eq:MOP} should include only those pairs of particles for which the line connecting the particle positions crosses the plane (see Figure \ref{fig:MOPdiag}a). We have verified that this method produces results in good agreement with global pressure measurements (for homogeneous systems) and with other methods for measuring the local 
pressure, even when the plane is very small, of the order of a few particle diameters. In our system, since there are no solvent-membrane interactions, the osmotic pressure is also given by the magnitude of the forces of confinement on the solute particles, per unit area of the confining boundary (as noted by Luo and Roux \cite{luo}). This provides an alternative way to measure the osmotic pressure, with results in good agreement with various other methods for determining the local pressure \cite{Lion2012}.  

To define the concentration of solute particles in our simulations, we need to define an effective boundary for the solution compartment. To do this, we match the pressure-density relation of a ``gas'' of solute particles, confined in the solution compartment in the absence of solvent (shown in Figure \ref{fig:MOPdiag}b, closed circles), with that of a periodic box of equivalent particles (Figure \ref{fig:MOPdiag}b, open circles). This results in an effective boundary of the solution compartment which lies $0.91_{-0.02}^{+0.07}\sigma$ inside the locus of divergence of the the confining potential. The error bars on the position of this boundary were computed by determining the range of boundary positions over which the simulated pressure-density curve  for the confined  solute ``gas'' can be said to match both the hard-sphere Carnahan-Starling prediction and simulation results for an unconfined (periodic) gas of WCA particles, with 95\% confidence (evaluated using a t-test).

\section{Hard sphere prediction for density imbalance}\label{app:CS}

In Figure \ref{fig:density2}, we compare our simulation results for the solvent density imbalance between solution and solvent compartments, to thermodynamic predictions for a system of hard spheres, obtained using the  Carnahan-Starling expression for the free energy of a system of hard spheres \cite{JPHansen,CarnahanStarling1969}. The solvent compartment, of volume $V^{\mathrm{out}}$, contains $N_{\mathrm{v}}^{\mathrm{out}}$ solvent particles, at packing fraction  $\eta_{\mathrm{v}}^{\mathrm{out}}$ (this is related to the density $\rho_{\mathrm{v}}^{\mathrm{out}} $ by $\eta_{\mathrm{v}}^{\mathrm{out}} = \pi \rho_{\mathrm{v}}^{\mathrm{out}} \sigma^3/6$). The free energy is then

\begin{eqnarray}
\frac{F^{\mathrm{out}}}{k_BT} &=& N_{\mathrm{v}}^{\mathrm{out}}\left[\log\left(\rho_{\mathrm{v}}^{\mathrm{out}} \Lambda^{3}\right)-1 + \eta_{\mathrm{v}}^{\mathrm{out}}\frac{(4-3\eta_{\mathrm{v}}^{\mathrm{out}})}{(1-\eta_{\mathrm{v}}^{\mathrm{out}})^{2}}\right]\\\nonumber &=& \frac{6\eta_{\mathrm{v}}^{\mathrm{out}} V^{\mathrm{out}}}{\pi\sigma^3}\left[\log\left(\frac{6 \eta_{\mathrm{v}}^{\mathrm{out}} \Lambda^{3}}{\pi \sigma^3}\right)-1 + \eta_{\mathrm{v}}^{\mathrm{out}}\frac{(4-3\eta_{\mathrm{v}}^{\mathrm{out}})}{(1-\eta_{\mathrm{v}}^{\mathrm{out}})^{2}}\right]
 \label{eq:HSfreeenergy}
\end{eqnarray}

where $\Lambda$ is the de Broglie thermal wavelength. The chemical potential is given by $\mu_{\mathrm{v}}^{\mathrm{out}}=\partial F^{\mathrm{out}}/\partial N_{\mathrm{v}}^{\mathrm{out}} = (\pi\sigma^3/6V^{\mathrm{out}})\partial F^{\mathrm{out}}/\partial \eta_{\mathrm{v}}^{\mathrm{out}}$: 

\begin{eqnarray}\hspace{-0.85cm}
\frac{\mu_{\mathrm{v}}^{\mathrm{out}}}{k_BT} &=& \log\left(\rho_{\mathrm{v}}^{\mathrm{out}} \Lambda^{3}\right)+ \eta_{\mathrm{v}}^{{\mathrm{out}}}\left[\frac{8-9\eta_{\mathrm{v}}^{{\mathrm{out}}}+3(\eta_{\mathrm{v}}^{{\mathrm{out}}})^{2}}{(1-\eta_{\mathrm{v}}^{{\mathrm{out}}})^{3}}\right]
 \label{eq:musolvent1}
\end{eqnarray}

The solution compartment, of volume $V^{\mathrm{in}}$, contains a mixture of solute particles, at packing fraction $\eta_{\mathrm{s}}$, and solvent particles, at packing fraction $\eta_{\mathrm{v}}^{\mathrm{in}}$. Its free energy is 

\begin{eqnarray}\hspace{-0.85cm}
\frac{F^{\mathrm{in}}}{k_BT} &=&  N_{\mathrm{s}}\left[\log\left(\rho_{\mathrm{s}} \Lambda^{3}\right)-1\right] +  N_{\mathrm{v}}^{\mathrm{in}}\left[\log\left(\rho_{\mathrm{v}}^{\mathrm{in}} \Lambda^{3}\right)-1 \right]\\\nonumber &+& \frac{6V^{\mathrm{in}}(\eta_{\mathrm{s}}+\eta_{\mathrm{v}}^{{\mathrm{in}}})^2 }{\pi\sigma^3}\left[\frac{4-3(\eta_{\mathrm{s}}+\eta_{\mathrm{v}}^{{\mathrm{in}}})}{1-(\eta_{\mathrm{s}}+\eta_{\mathrm{v}}^{{\mathrm{in}}})^{2}}\right]
 \label{eq:HSfreeenergy2}
\end{eqnarray}

The chemical potential $\mu_{\mathrm{v}}^{\mathrm{in}} = \partial F^{\mathrm{in}}/\partial N_{\mathrm{v}}^{\mathrm{in}} = (\pi\sigma^3/6V^{\mathrm{in}})\partial F^{\mathrm{in}}/\partial \eta_{\mathrm{v}}^{\mathrm{in}}$: 

\begin{eqnarray}
\frac{\mu_{\mathrm{v}}^{\mathrm{in}}}{k_BT} &=& \log\left(\rho_{\mathrm{v}}^{\mathrm{in}} \Lambda^{3}\right)\\\nonumber &+& (\eta_{\mathrm{s}}+\eta_{\mathrm{v}}^{{\mathrm{in}}})\left[\frac{8-9(\eta_{\mathrm{s}}+\eta_{\mathrm{v}}^{{\mathrm{in}}})+3(\eta_{\mathrm{s}}+\eta_{\mathrm{v}}^{{\mathrm{in}}})^{2}}{(1-(\eta_{\mathrm{s}}+\eta_{\mathrm{v}}^{{\mathrm{in}}}))^{3}}\right].
 \label{eq:musolvent2}
\end{eqnarray}

Setting $\mu_{\mathrm{v}}^{\mathrm{in}}=\mu_{\mathrm{v}}^{\mathrm{out}}$, we arrive at the following relation:

\begin{eqnarray}
\nonumber \log\left(\frac{\eta_{\mathrm{v}}^{\mathrm{in}}}{\eta_{\mathrm{v}}^{\mathrm{out}}}\right)& +&  (\eta_{\mathrm{s}}+\eta_{\mathrm{v}}^{{\mathrm{in}}})\left[\frac{8-9(\eta_{\mathrm{s}}+\eta_{\mathrm{v}}^{{\mathrm{in}}})+3(\eta_{\mathrm{s}}+\eta_{\mathrm{v}}^{{\mathrm{in}}})^{2}}{(1-(\eta_{\mathrm{s}}+\eta_{\mathrm{v}}^{{\mathrm{in}}}))^{3}}\right]\\ &= & \eta_{\mathrm{v}}^{{\mathrm{out}}}\left[\frac{8-9\eta_{\mathrm{v}}^{{\mathrm{out}}}+3(\eta_{\mathrm{v}}^{{\mathrm{out}}})^{2}}{(1-\eta_{\mathrm{v}}^{{\mathrm{out}}})^{3}}\right].
 \label{eq:densitydiff}
\end{eqnarray}

Conservation of solvent particle number implies that 

\begin{equation}
V^{\mathrm{in}}\eta_{\mathrm{v}}^{\mathrm{in}} + V^{\mathrm{out}}\eta_{\mathrm{v}}^{\mathrm{out}} = \frac{\pi \sigma^3 N_{\mathrm{v}}}{6} \label{eq:cons}
\end{equation}

where $ N_{\mathrm{v}}$ is the total number of solvent particles. Eqs.~(\ref{eq:densitydiff}) and (\ref{eq:cons}) are solved numerically to obtain $\eta_{\mathrm{v}}^{\mathrm{in}}$ and $\eta_{\mathrm{v}}^{\mathrm{out}}$ as functions of the solute packing fraction  $\eta_{\mathrm{s}}$, as plotted in Figure \ref{fig:density2}.

\section{Derivation of Eq.(\ref{eq:vHpluscorr}) for  the osmotic pressure}\label{app:virial}

Eq.~(\ref{eq:vHpluscorr}) follows from the dynamical equations of motion for the solute particles, under the conditions that the system is in steady state and the solute particles are confined in the solution compartment. The first of these conditions implies that  $d \left\langle \sum_{\mathrm{i_s}} \vec{r_i}\cdot \vec{p_i} \right\rangle/dt = 0$ (where the sum is over all solute particles and $ \vec{r_i}$ and $ \vec{p_i} $ are the position and momentum of particle $i$ respectively). Expanding the time derivative and using Newton's equations of motion leads to: 

\vspace{-0.35cm}
\begin{equation}\label{eq:clausius}
\left \langle  \sum_{\mathrm{i_s}} \frac{|\vec{p_i}|^2}{m_i}\right \rangle + \left\langle \sum_{\mathrm{i_s}} \vec{r_i}\cdot \vec{F_i} \right \rangle = 0
\end{equation}
\vspace{-0.35cm}

where $\vec{F_i}$ is the total force acting on solute particle $i$, and $m_i$ is its mass. Eq.(\ref{eq:clausius}) constitutes a Clausius virial relation for the solute particles \cite{AllenTildesley,JPHansen,Lion2012}. Splitting the force $\vec{F_i}$ into the contributions due to interactions with the confining potential, solute-solute interactions and solute-solvent interactions, and using the equipartition theorem, we arrive at: 

\begin{equation}
 3N_{\mathrm{s}}k_BT + \sum_{\mathrm{i_s}} \sum_{\mathrm{j_s>i_s}}\vec r_{i}.\vec f_{ij} +  \sum_{\mathrm{i_s}}\sum_{\mathrm{j_v}} \vec r_{i}.\vec f_{ij} +  \sum_{\mathrm{i_s}} \vec r_{i}.\vec{f}_{i,\mathrm{conf}} = 0,
 \label{eq:solutevirial}
\end{equation}

where  $\vec f_{ij}$ is the force exerted on particle $i$ by particle $j$  and $\vec{f}_{i,\mathrm{conf}}$ is the confining force on particle $i$. The first sum is over solute-solute pairs (without double counting), the second sum is over solute-solvent pairs, and the final sum is over interactions between solute particles and the ``membrane''. Focusing on the final term, we note that, to a good approximation, solute particles only feel the confining force when they are very close to the boundary of the solution compartment. The final term can then be approximated by an  integral over a narrow shell at the boundary of the solution compartment. Taking into account also the fact that the osmotic pressure is given by the average normal force per unit area on the membrane, which in our system is equivalent to the magnitude of the confining force on the solute particles, per unit area, leads to \cite{Lion2012}

\vspace{-0.4cm}
\begin{equation}\label{eq:os}
\sum_{\mathrm{i_s}} \vec r_{i}.\vec{f}_{i,\mathrm{conf}} \approx -3V^{\mathrm{in}} \Delta P
\end{equation}
\vspace{-0.4cm}

where $\Delta P$ is the osmotic pressure and $V^{\mathrm{in}}$ is the volume of the solution compartment. Substituting Eq.(\ref{eq:os}) into Eq.(\ref{eq:solutevirial}) leads directly to Eq.(\ref{eq:vHpluscorr}) in the main text.

\end{appendix}

\end{document}